\documentstyle[12pt,a4,epsfig]{article}

\newlength{\abstwidth}
\begin{document}
\newcommand{\hi }{{\footnotesize HIJING}}
\newcommand{\hij}{{\footnotesize HIJING }}
\newcommand{\qgp}{{\footnotesize QGP }}
\newcommand{\qgpp}{{\footnotesize QGP}}
\newcommand{\qcd}{{\footnotesize QCD }}
\newcommand{\pqcd}{{\footnotesize pQCD }}
\newcommand{\lhc}{{\footnotesize LHC }}
\newcommand{\rhic}{{\footnotesize RHIC }}
\newcommand{\sfm}{{\footnotesize SFMs }}
\newcommand{\flt}{{fluctuations }}
\newcommand{\fltt}{{fluctuations}}
\newcommand{\nudyn}{$\nu_{[+-,dyn]}$ }
\newcommand{\nudy}{$\nu_{dyn}$}
\newcommand{\nudync}{$\nu_{[+-,dyn]}^{corr}$ }
\newcommand{\cor}{correlations }
\newcommand{\urqmd}{{\footnotesize URQMD }}

\thispagestyle{empty}
\begin{center}
{\Large {\bf Contributions of Jets in Net Charge Fluctuations from the Beam Energy Scan at RHIC and LHC}}\\ [5mm]

	{\bf Bushra Ali, Shaista Khan and Shakeel Ahmad\footnote{email: Shakeel.Ahmad@cern.ch\\Authors declare that there is no conflict of interest}}\\
{\it  Department of Physics, Aligarh Muslim University\\ [-2mm] Aligarh - 202002 INDIA}

\end{center}

\begin{center}
{\bf Abstract}\\ [2ex]
\begin{minipage}{\abstwidth} Dynamical net charge \flt have been studied in ultra-relativistic heavy-ion collisions from the beam energy scan at \rhic and \lhc energies by carrying out the hadronic model simulation. Monte Carlo model, \hij is used to generate events in two different modes, \hi -default with jet quenching switched off and jet/minijet production switched off. A popular variable, \nudyn is used to study the net charge \flt in different centrality bins and the findings are compared with the available experimental values reported earlier. Although the broad features of net charge \flt are reproduced by the \hi, yet the model predicts the larger magnitude of \flt as compared to the one observed in experiments. The role of jets/minijets production in reducing the net charge \flt is, however distinctly visible from the analysis of the two types of \hij events. Furthermore, $dN_{ch}/d\eta$ and $1/N$ scaling is partially exhibited which is due to the fact that in \hi, nucleus-nucleus collisions are treated as multiple independent nucleon-nucleon collisions. \\ \\

[10ex]
\end{minipage}
\end{center}

%\begin{itemize}
\noindent KEY-WORDS: Beam Energy Scan, Net Charge Fluctuations, Relativistic heavy-ion collisions.\\
%\end{itemize}
\phantom{dummy}

%=================================================================================================
\newpage
\noindent {\bf 1. Introduction}\\

\noindent The interest in the studies involving event-by-event \flt in hadronic (hh) and heavy-ion (AA) collisions is primarily connected to the idea that the \cor and \flt of dynamical origin are associated with the critical phenomena of phase transitions and leads to the local and global differences between the events produced under similar initial conditions [1,2]. Several different approaches have been made to investigate the event-by-event fluctuations in hh and AA collisions at widely different energies, e.g., multifractals [3,4,5], normalized factorial moments [6], erraticity [4,7], k-order pseudorapidity spacing [8,9] and transverse momentum($p_T$) spectra, etc. Furthermore, event-by-event \flt in the conserved quantities, like, strangeness, baryon number, electric charge have emerged as new tools to estimate the degree of equilibration and criticality of the measured system [12]. Experiments such as \rhic and \lhc are well suited for the study of these observables [12,13].\\

\noindent Event-by-event \flt of net charge of the produced relativistic charged particles serve as an important tool to investigate the composition of hot and dense matter prevailing in the `fireball', created during the intermediate stage of AA collisions, which, in principle, be characterized in the framework of \qcd  [13]. It has been argued that a phase transition from \qgp to normal hadronic state is an entropy conserving process [14] and therefore, the \flt in net electric charge will be significantly reduced in the final state in comparison to what is envisaged to be observed from a hadron gas system [15,18]. This is expected because the magnitude of charge \flt is proportional to the square of the number of charges present in the system which depends on the state from which charges originate. A system passing through \qgp phase, quarks are the charge carriers whereas in the case of hadron gas the charge carriers are hadrons. This suggests that the charge \flt observed in the case of \qgp with fractional charges would be smaller than those in hadron gas with integral charges [12,16,19]. A reduction in the \flt of net charge in Pb-Pb collision at $\sqrt{s_{NN}} = 2.76$ TeV in comparison to that observed at \rhic has been reported by {\footnotesize ALICE} collaboration [20]. A question arises here whether the \flt arising from \qgp or from hadron gas would survive during the evaluation of the system [12,21,22,23,24]. The \flt observed at the freeze-out depend crucially on the equation of state of the system and final effects. It has been shown [25] that large charge \flt survive, if accompanied by large temperature fluctuations at freeze-out in context to the experiments. Measurement of charge \flt depends on the observation window which is so selected that the majority of the \flt are captured without being affected by the conservation limits [22,23,24].\\

\noindent An attempt is, therefore, made to carry out a systematic study of dynamical net charge \flt from beam energy scan at \rhic and \lhc energies using the Monte-Carlo model, \hij and the findings are compared with those obtained with the real data and other MC models. The reason for using the code \hij is that it gives an opportunity to study the effect of jets and jet-quenching. \hij events are generated at various beam energies corresponding to \rhic and \lhc which covers an energy range from $\sqrt{s_{NN}} =$ 62.4 GeV to 5.02 TeV. Two sets of events, i) \hi -default with jets and minijets and ii) \hij with no jet/minijet production are generated for each of the incident energies considered. \\

%=========================================================================================
\noindent {\bf 2. Formalism}\\

\noindent The charge \flt are usually studied in terms of two types of measures [26]. The first one is the D which is the direct measure of the variance of event-by-event net charge $\langle \delta Q^2 \rangle = \langle Q^2 \rangle - \langle Q \rangle^2$, where $Q = N_+ - N_-$; $N_+$ and $N_-$ respectively denote the multiplicities of positively and negatively charged particles produced in an event in the considered phase space. Since the net charge \flt may get affected by the uncertainties arising out of volume \fltt, the \flt in the ratio, $R = \frac{N_+}{N_-}$ is taken as the other suitable parameter. R is related to the net charge \flt via the parameter D as [15,16,19,20]:
\begin{eqnarray}
              D = \langle N_{ch} \rangle \delta R^2 \simeq 4\frac{\langle \delta Q^2 \rangle}{\langle N_{ch} \rangle}
\end{eqnarray}
which gives a measure of charge \flt per unit entropy. It has been shown that D acquires a value $\sim 4$ for an uncorrelated pion gas which decreases to $\sim 3$ after taking into account the resonance yields [16]. For \qgpp, the value of D has been reduced to $\sim 1 - 1.5$, where the uncertainty arises due to the uncertainties involved in relating the entropy to the multiplicity of the charged hadrons in the final state [27]. The parameter D, thus, may be taken as an efficient probe for distinguishing between the hadron gas and \qgp phases. These \flt are however, envisaged to be diluted in the rapidly expanding medium due to the diffusion of particles in rapidity space [22,23]. Resonance decays, collision dynamics, radial flow and final state interactions may also affect the amount of \flt measured [16,28,29,30]. The first results on net charge \flt at \rhic were presented by {\footnotesize PHENIX} [31] in terms of reduced variance $\omega_d = \frac{\langle \delta Q^2 \rangle}{N_{ch}}$ while {\footnotesize STAR} [30] results were based on a dynamical net charge \flt measure, \nudyn and were treated as a rather reliable measure of the net charge \flt as \nudyn was found to be robust against detection efficiency. \\

\noindent Furthermore, the contributions from statistical \flt would also be present if net charge \flt are studied in terms of parameter D and it will be difficult to extract the contribution due to \flt of dynamical origin. The novel method of estimating the net charge \flt takes into account the correlation strength between +\ +, -\ -, and +\ - charge particle pairs [12,32]. The difference between the relative multiplicities of positively and negatively charged particles is given as,
\begin{eqnarray}
\nu_{+-} = \langle (\frac{N_+}{\langle N_+ \rangle}- \frac{N_-}{\langle N_- \rangle})^2 \rangle
\end{eqnarray}
where the angular brackets represent the mean value over the entire sample of events. The Poisson limit of this quantity is expressed as [30]:
\begin{eqnarray}
\nu_{[+-,stat]} = \frac{1}{\langle N_+ \rangle}+ \frac{1}{\langle N_- \rangle}
\end{eqnarray}
The dynamical net charge \flt may, therefore, be written as the difference of these two quantities:
\begin{eqnarray}
 {\Large \nu_{[+-,dyn]} = \nu_{[+-]} - \nu_{[+-,stat]} }
\end{eqnarray}
\begin{eqnarray}
\nu_{[+-,dyn]} = \frac{\langle N_+(N_+ - 1)\rangle}{\langle N_+\rangle^2} + \frac{\langle N_-(N_- - 1)\rangle}{\langle N_-\rangle^2} -2\frac{\langle N_+N_-\rangle}{\langle N_+\rangle\langle N_-\rangle}
\end{eqnarray}
From the theoretical point of view, \nudyn can be expressed in terms of two particle integral correlation functions as
\begin{eqnarray}
\nu_{[+-,dyn]} = R_{++} + R_{--} - 2R_{+-}
\end{eqnarray}
where the term $R_{\alpha\beta}$  gives the ratio of integrals of two- and single-particle pseudorapidity density function, defined as,
\begin{equation}
R_{\alpha\beta} = \frac{\int dn_\alpha dn_\beta \frac{dN}{dn_\alpha dn_\beta}}{\int dn_\alpha \frac{dN}{dn_\alpha} \int dn_\beta \frac{dN}{dn_\beta}}
\end{equation}
\noindent The variable, \nudyn is, thus, basically a measure of relative correlation strength of +\ +, -\ -, and +\ - charged hadron pairs. For independent emission of particles, these \cor should be ideally zero. However, in practice, a partial \cor is observed due to string and jet-fragmentation, resonance decays, etc. The strength of $R_{++}$, $R_{--}$ and $R_{+-}$ are expected to vary with system size and beam energy. Moreover, as the charge conservation, +\ - pair are expected to be rather strongly correlated as compared to like sign charge pairs and hence $2R_{+-}$ in Eq.6 is envisaged to be larger than the sum of the other two terms [30] giving \nudyn values less than zero, which is evident from the results based on $pp$ and $\bar{p}p$ collisions at {\footnotesize CERN ISR, FNAL} and later on in heavy-ion collisions at \rhic[30,32,34,35,36] and \lhc energies [20,30].\\
%=========================================================================================

\noindent{\bf 3. Results and discussion}\\

\noindent Several sets of MC events corresponding to different collision systems in a wide range of beam energies are generated using the code \hi - 1.37 [37] for the present analysis. The details of the events simulated are listed in Table-1. Two sets of events for each beam energy and colliding nuclei, \hi-default with jet-quenching off and with jet/minijet production switched off are simulated and analysed. It has been argued [38,39] that the minijets (semi-hard parton scattering with few GeV/c momentum transfer) are copiously produced in the early state of AA collisions at \rhic and higher energies. In a \qgp medium, if present, the jets/minijets will lose energy through induced gluon radiation [40], a process referred to as jet quenching in the case of higher $p_{T}$ partons. The properties of the dissipative medium would determine the extent of energy loss of jets and minijets. The influence of the production of jets/minijets in AA collisions in the produced medium on the net charged \flt may be investigated by comparing the findings due to the two types of \hij simulated events. The analysis has been carried out by considering the particles having their pseudorapidity values $|\eta| < 1.0$ and $p_{T}$ values in the range 0.2 GeV/c $< p_T <$ 5.0 GeV/c. These $\eta$ and $p_T$ cuts have been applied to facilitate the comparison of the findings with the experimental result having similar cuts. \\

\noindent Values of \nudyn for different collision centralities are estimated for various data sets and are listed in Tables 2 - 4 along with the corresponding values of number of participating nucleons, $N_{part}$. Variation of \nudyn with mean number of participating nucleons, $N_{part}$, for various data sets are exhibited in Fig.1. Such dependences observed in experiments, {\footnotesize STAR} [30] and {\footnotesize ALICE} [20,36] are also displayed in the same figure. A monotonic dependence of \nudyn on $N_{part}$ is seen in the figure. It may be of interest to note that for a given $N_{part}$, the magnitude of \nudyn decreases with increasing beam energy and this difference becomes more and more pronounced on moving from most central ($5\% $) to the peripheral ($70 - 80\% $) collisions. It is also interesting to note in the figure that the \hij predicted values (for \hi -default events) are quite close to the experimental values. However, the corresponding \nudyn values for the events with jets/minijets off are somewhat larger. The jets-off multiplicities reflects the soft processes, whereas, the jets-on multiplicities includes the contributions from the jets and minijets [36]. This may cause the reduction in the contributions coming from the third term of Eq.5 which represents the correlations between +\ - pairs. This is expected to occur at these energies, as the events have high multiplicities and are dominated by multiple minijet production which might cause the reduction in the strengths of correlations and \flt [41]. \\

\noindent The parameter D and \nudyn are related to each other as per the relation: 
\begin{eqnarray}
\langle N_{ch}\rangle\nu_{[+-,dyn]} = D - 4 
\end{eqnarray}
\noindent The magnitude of net charge \flt is limited by the global charge conservation of the produced particles [32]. Considering the effect of global charge conservation, the dynamical \flt need to be corrected by a factor of $-4/\langle N_{total}\rangle$, where $N_{total}$ denotes the total charged particle multiplicity of an event in full phase space. Taking into account the global charge conservation and finite acceptance, the corrected value of \nudyn is given by,
\begin{eqnarray}
	\nu_{[+-,dyn]}^{corr} =  \nu_{[+-,dyn]} + \frac{4}{\langle N_{total}\rangle}
\end{eqnarray}
\noindent Values of \nudync for various data sets are presented in the last column of Tables 2 - 4, whereas variations of \nudync with $N_{part}$ for these data sets are displayed in Fig.2. Although the trends of variations of \nudyn and \nudync with $N_{part}$ for both types of \hij events are similar, yet it might be noticed that the data points corresponding to various energies lie rather close to each other in the semi-central and peripheral collision regions. This weakening of energy dependence is observed for both types of \hij samples considered. \\

\noindent The observed dependence of \nudyn or its corrected form \nudync, on $N_{part}$ or collision centrality indicates the weakening of correlations amongst the produced hadrons, as one moves from central to peripheral collisions, and nearly match with the experimental results. These findings, thus, tend to suggest that the \nudyn should be proportional to the centrality of collisions or charged particle multiplicity, if AA collisions are taken as the superpositions of independent nucleon-nucleon (nn) collisions with negligible re-scattering effects (which is the basic property of \hij model). This may be tested by scaling the \nudyn by charged particle density $dN_{ch}/d\eta$, and plotting against $N_{part}$. These plots are displayed in Figs.3 and 4. It may be observed from these figures that the data at different energies show the same qualitative behavior. The values of product $(dN_{ch}/d\eta)$\nudyn are noticed to be minimum for peripheral collisions and gradually increase to their maximum for the most central collisions; the rise from minimum to maximum is about $\sim $ 35 - 40 $\%$ for various data sets. An increase of $50\%$ has been observed [40] in {\footnotesize STAR} Au-Au collisions. Such an increase in $(dN_{ch}/d\eta)$\nudyn values with $N_{part}$ may be accounted due to the increase in the particle multiplicity per participant. Data from {\footnotesize UA1} and {\footnotesize PHOBOS} show that for pp and Au-Au collisions at 200 GeV, $dN_{ch}/d\eta$ increases from 2.4 to 3.9 for most central collisions, thus giving an increase of about $60\%$ [42] \\

\noindent The scaling of \nudyn with $N_{part}$ has also been checked and the plots are shown in Fig.5, whereas after applying the corrections to \nudyn the values of the products are plotted against $N_{part}$ in Fig.6. It is observed from these figures that with increasing $N_{part}$, $N_{part}$\nudyn values gradually decrease for all the data sets. Moreover, for a given $N_{part}$ the values of product $N_{part}$\nudyn decrease with the beam energy. It is interesting to note that the difference in the values observed at \rhic and \lhc energies, after applying the corrections to \nudyn values almost vanishes. It is also interesting to note that the \hij simulated data points lie closer to the corresponding ones reported earlier using the {\footnotesize ALICE} data [20]. The decreasing trends of $N_{part}$\nudyn (or $N_{part}$\nudync) from peripheral to most central collisions observed in {\footnotesize STAR} are in contrast to what is observed in the present study using the \hij data at \rhic and higher energies. Furthermore, the lower values of product $(dN_{ch}/d\eta)$\nudyn or $N_{part}$\nudyn, as shown in Figs.4 and 6 predicted by the \hij with no jets in comparison to those predicted by \hi -default, indicates the reduction in magnitude of \nudyn due to the productions of jets and minijets. \\

\noindent The variations of \nudyn and \nudync with charged particle density, $dN_{ch}/d\eta$ for the two sets of \hij events are shown in Fig.7. Results based on Pb-Pb 2.76 TeV experimental data [20] for the same $\eta$ and $p_T$ cuts are also presented in the same figure. It is worth while to note in these figures that \hi -default predicted values for 2.76 TeV data are quite close to the corresponding experimental values. Although the magnitude of \nudyn or \nudync exhibits an energy dependence, which becomes more pronounced as the $dN_{ch}/d\eta$ values decrease, i.e., from semi-central to peripheral collisions, yet the data points for various event samples tend to fall on a single curve. Data for the events with no jets exhibit almost similar behaviour except for Pb-Pb data at 2.76 and 5.02 TeV without jet production. This may lead to the conclusion that as one moves from \rhic to \lhc energies, contributions to the particle multiplicity coming from the jet/minijet production causes the reduction in the magnitude of charge \fltt. \\

\noindent As mentioned earlier, if AA collisions are the superpositions of m number of nn collisions the single particle density for nn and AA collisions would be written as: $\rho_1^nn(\eta) = dN_{ch}/d\eta$ and $\rho_1^AA(\eta) = m\rho_1^{nn}{\eta}$. In such a scenario, the invariant cross section is proportional to the number of nn collisions, m, and the quantity $(dN_{ch}/d\eta)$\nudyn is independent of  centrality of collision and the system size [14]. {\footnotesize STAR} results, however, give $\sim 40\%$ increase in $(dN_{ch}/d\eta)$\nudyn values for Au-Au and Cu-Cu collisions. The product $(dN_{ch}/d\eta)$\nudyn is plotted against $dN_{ch}/d\eta$ for the two types of event sample in Fig.8. Similar plots for \nudync are also shown in Fig.9. The scaled values of \nudyn and \nudync are observed to increase with increasing $dN_{ch}/d\eta$ values in almost similar fashion. Furthermore, for a given $dN_{ch}/d\eta$ the scaled values of \nudyn or its corrected version are noticed to increase with increasing energy. It is also observed that for a particular set of events(\hi -default and jets off) the values of \nudyn and \nudync are somewhat larger when jet/minijet production is switched off. \\

\noindent It has been suggested [43] that any multiplicity scaling should be based on the mean multiplicities of charged particles. In the model independent sources [44], mean particle multiplicity is taken to be proportional to the number of sources, $\langle N_s \rangle$, which changes from event to event. The multiplicity of positively and negatively charged particles may be expressed as
\begin{eqnarray}
\langle N_+\rangle = \alpha_1+\alpha_2+..... +\alpha_{N_s} \\
\langle N_-\rangle = \beta_1+\beta_2+....+\beta_{N_s}
\end{eqnarray} 
where, $\alpha_i$ and $\beta_i$ represent the contributions from $i^{th}$ source. The first and second moments of multiplicity distributions are written as
\begin{eqnarray}
\langle N_a\rangle = \langle \alpha \rangle\langle N_s\rangle \\
\langle N_b\rangle = \langle \beta\rangle\langle N_s\rangle \\
\langle N_a^2 \rangle = \langle \alpha^2\rangle\langle N_s\rangle+\langle \alpha\rangle^2[\langle N_s^2\rangle-\langle N_s\rangle] \\
\langle N_b^2 \rangle = \langle \beta^2\rangle\langle N_s\rangle+\langle \beta\rangle^2[\langle N_s^2\rangle-\langle N_s\rangle] \\
\langle N_aN_b \rangle = \langle \alpha\beta\rangle\langle N_s\rangle+\langle \alpha\rangle\langle \beta\rangle\langle N_s^2\rangle-\langle N_s\rangle]
\end{eqnarray}
here $\langle \alpha\rangle$, $\langle \beta\rangle$ and $\langle \alpha^1\rangle$, $\langle \beta^1\rangle$, $\langle \alpha\beta\rangle$ are the first and second moments of the probability distributions $P(\alpha,\beta)$ for a single source. \\
\noindent Following the details as given in ref [44] and using the equation:
\begin{eqnarray}
\nu_{dyn}[a,b] = \frac{\langle N_a^2\rangle}{\langle N_a \rangle^2} + \frac{\langle N_b^2\rangle}{\langle N_b\rangle^2} - 2\frac{\langle N_aN_b\rangle}{\langle N_a\rangle \langle N_b\rangle} - (\frac{1}{\langle N_a \rangle} + \frac{1}{\langle N_b \rangle})
\end{eqnarray}
the following form of \nudy may be obtained [45]
\begin{eqnarray}
\nu_{dyn}[a,b] = \frac{1}{\langle N_s\rangle}[\frac{\langle \alpha^2\rangle}{\langle \alpha\rangle^2}+ \frac{\langle \beta^2\rangle}{\langle \beta\rangle^2} - 2\frac{\langle \alpha\beta\rangle}{\langle \alpha\rangle\langle \beta\rangle} - (\frac{1}{\langle \alpha\rangle}+\frac{1}{\langle \beta\rangle})]
			\simeq \frac{1}{\langle N_s\rangle}\nu^*[\alpha,\beta]
\end{eqnarray}
\noindent where, $\nu^*[\alpha,\beta]$ is the quantity of the multiplicities of types a and b for each source. This gives $\nu_{a,b}$ to be inversely proportional to the size of the colliding nuclei. On the other hand, as the term $\langle N_s^2\rangle - \langle N_s\rangle$ cancels out by construction, \nudy is independent of the system size but requires an additional scaling due to the remaining term, $1/\langle N_s\rangle$. If $1/(\frac{1}{\langle N_a\rangle}+\frac{1}{\langle N_b\rangle})$ type of scaling is used, then substituting Eqs.12 and 13 in Eq.17, the term $1/\langle N_s\rangle$ vanishes and the following form of the scaling is obtained:
\begin{eqnarray}
	\frac{\nu_{dyn}[a,b]}{\frac{1}{\langle N_a\rangle}+\frac{1}{\langle N_b\rangle}} = \frac{\nu_{dyn}[\alpha,\beta]}{\frac{1}{\langle \alpha\rangle}+\frac{1}{\langle \beta\rangle}}
\end{eqnarray}
\noindent The scaling of this type has been tested and the results for the various data sets are shown in Figs.10 and 11. It may be seen in these figures that the scaled \nudyn values for a given energy are nearly independent of charged particle density. It is, further, observed that the magnitude of scaled \nudyn values increases as one moves from \rhic to \lhc energies. The magnitude of \nudyn is observed to be inversely proportional to the number of sub collisions leading to the particle production. If number of particles produced in each sub collisions is independent of collision centrality, \nudyn would exhibit $1/N$ scaling [11]. It has been reported [11] that in Au-Au collisions at 130 GeV $1/N$ scaling is clearly noted by the data. \hij simulated data, however, supports such scaling. In contrast to this, findings from \urqmd simulations do not support $1/N$ scaling which maybe because in \urqmd re-scattering effects are included which would reduce the magnitude of $N$\nudyn for central collisions [11]. On the basis of various types of scaling of \nudyn tested in the present study and also the ones by other workers it may be concluded here that $1/(\frac{1}{\langle N_a\rangle}+\frac{1}{\langle N_b\rangle})$ scaling of \nudyn is relatively a better scaling as compared to other scalings.  \\ 
%===========================================================================

\noindent{\bf 4. Conclusions}\\

\noindent A systematic study of various aspects of net charge \flt has been looked into by simulating the Monte Carlo events using the \hij generator in two different modes, i) \hi -default with jet-quenching turned off and, ii) production of jets and minijets turned off. Although both types of events exhibit almost similar dependence of \nudyn on collision centrality and charged particle density, yet the observed difference in the magnitude of \flt clearly reflects the role of jets and minijets in reduction of net charge \fltt. The trend of energy dependence of \nudy, for various centrality bins, exhibited by the MC data used in the present study, match with {\footnotesize STAR} and {\footnotesize ALICE} results. $N_{part}$ and $dN_{ch}/d\eta$ scalings of \nudyn after applying the correction for global charge conservation are approximately exhibited by both types of event samples used. This is expected as in \hij case, AA collisions are treated as the superpositions of multiple nucleon-nucleon collisions. The findings also reveal that the production of jets and minijets plays dominant role in reducing the strength of particle \cor and \fltt.\\
%=========================================================================

\newpage
\noindent  {{\bf References}}
\begin{enumerate}
	\item [1] E. A. De Wolf et al., {\it Phys. Rept.} 270 (1996) 1.
	\item [2] Shakeel Ahmad et al, {\it Adv. in High En. Phys.} 2018 (2018) 6914627. 
	\item [3] R. C. Hwa, {\it Phys. Rev.} D41 (1990) 1456.
	\item [4] Shakeel Ahmad et al., {\it Chaos Solitons and Fractals}, 42 (2009) 538. 
	\item [5] Shaista Khan and Shakeel Ahmad, {\it Int. J. Mod. Phys.} E27 No. 1 (2018) 1850004.
	\item [6] A. Bialas and R. Peschanski, {\it Nucl. Phys.} B273 (1986) 703.
	\item [7] R. C. Hwa, {\it Acta. Phys. Polon.} 27 (1996) 1789.
	\item [8] M. L. Cherry et al., {\it Acta Phys. Polon.} B29 (1998) 2129.
	\item [9] K. Fialkowski and R. Wit, {\it Acta Phys. Pol.} B30 (1999) 2759.
	\item [10] J. C. Rider et al., {\it Nucl. Phys.} A698 (2002) 1998.
	\item [11] Claude A. Pruneau,({\footnotesize STAR} coll) {\it  Acta Phys. Hung.} A21 (2004) 261.
	\item [12] Bhanu Sharma et al, {\it Phys. Rev.} C91 (2015) 024909.
	\item [13] M. Mukherjee et al., arXiv: 1603.0283v3[nucl-ex].
	\item [14] S. Gosh et al, {\it Phys. Rev.} C96 (2017) 024912.
	\item [15] S. Jeon and V. Koch, {\it Phys. Rev. Lett.} 83 (1999) 5435.
	\item [16] S. Jeon and V. koch, {\it Phys. Rev. Lett.} 83 (2000) 2076.
	\item [17] M. Asakawa et al, {\it Phys. Rev. Lett.} 85 (2000) 2072.
	\item [18] H. Heiselberg and A. D. Jackson, {\it Phys. Rev.} C63 (2001) 064904.
	\item [19] M. Bleicher et al, {\it Phys. Rev.} C62 (2000) 061902.
	\item [20] B. Abelev et al, ({\footnotesize ALICE} coll) {\it Phys. Rev. Lett.} 110 (2013) 1523201.
	\item [21] B. Mohanty, J. Alam and T. K. Nayak, {\it Phys. Rev.} C67 (2003) 024904.
	\item [22] E. V. Shuryak and M. A. stephanov, {\it Phys. Rev.} C63 (2001) 064903.
	\item [23] M. Abdul Aziz and S. Gavin, {\it Phys. Rev.} C70 (2004) 034905.
	\item [24] M. Sakaida et al, {\it Phys. Rev.} C90 (2014) 064911.
	\item [25] M. Prakash, {\it Phys. Rev.} C65 (2002) 034906.
	\item [26] M. R. Atayan et al, {\it Phys. Rev.} D71 (2005) 012002.
	\item [27] S. Jeon and V. Koch in Quark-Gluon Plasma3, edited by R. C. Hwa and X. N. Wang, (World Scientific, Singapore 2004) p.430 arXiv: hep-ph/0304012v1.
	\item [28] S. Voloskin, {\it Phys. Lett.} B632 (2006) 024905.
	\item [29] J. Zarnek, {\it Phys. Rev.} C66 (2002) 024905.
	\item [30] B. J. Abelev et al,({\footnotesize STAR} coll) {\it Phys. Rev.} C79 (2009) 024906.
	\item [31] K. Adox et al,({\footnotesize PHENIX} coll) {\it Phys. Rev. Lett.} 89 (2002) 082301.
	\item [32] Claude A. Pruneau et al, {\it  Phys. Rev.} C66 (2002) 0444904.
	\item [33] J. Whitmore, {\it Phys. Rev.} 27 (1976) 187.
	\item [34] L. Foa, {\it Phys. Rev.} 22 (1975) 1.
	\item [35] J. Adamus et al,({\footnotesize STAR} coll) {\it Phys. Rev.} C68 (2003) 044905.
	\item [36] Zhou You et al,{\it Chin. Phys.} C34 (2018) 1436.
	\item [37] M. Gyulassy and Xin-Nian Wang, {\it Comput. Phys. Commun.} G25 (1999) 1859.
	\item [38] Q. Liu and T. A. Trainor, arXiv: hep-ph/031214v1.
	\item [39] K. Kajantie et al, {\it Phys. Rev. Lett.} 59 (1987) 2527.
	\item [40] M. Gyulassy and M. Plumer, {\it Phys. Rev.} B243 (1990) 432.
	\item [41] X. N. Wang, {\it Phys. Rep.} B248 (1990) 447.
	\item [42] Claude A. Pruneau,({\footnotesize STAR} coll) {\it Acta Phys. Hung.} A25 (2006) 401.
	\item [43] V. Koch and T. Schuster, {\it Phys. Rev.} C81 (2010) 034910.
	\item [44] A. Bialas et al, {\it Nucl. Phys.} B111 (1976) 46.
	\item [45] M. Arslandok, {\it Nucl. Phys.} A956 (2016) 870.	

\end{enumerate}

%================= Table-1 =============================================

\newpage
\begin{table}
\centering
\caption{Details of events selected for analysis.}\vspace{2mm}
\begin{tabular}{c|c|c} \hline 
Energy & Type of    & No. of events\\
(GeV)& collision  & ($\times 10^{6}$) \\  [2mm] \hline
5020   &    Pb-Pb   &  0.6 \\   
2760   &    Pb-Pb   &  0.6 \\  
200   &    Au-Au   &  0.6 \\ 
130   &    Au-Au   &  0.6 \\ 
100   &    Au-Au   &  0.6 \\ 
200   &    Cu-Cu   &  1.0 \\ 
62.4    &    Cu-Cu   &  1.0 \\ \hline 
\end{tabular}
\end{table}

%==============================Table-2==================================

\newpage
\begin{table}
\footnotesize
\centering
\caption{Values of $N_{part}$, $\nu_{[+-,dyn]}$ and $\nu_{[+-,dyn]}^{corr}$ for various centrality classes in $|\eta| < 1$ for events corresponding $^{197}Au - ^{197}Au$ collisions.}\vspace{2mm}
\begin{tabular}{c|rrr|rrr} \hline \hline
 &\multicolumn{3}{c|}{HIJING - default}  & \multicolumn{3}{c}{HIJING - nojets} \\
cent.$\%$ & $N_{part}$  &    $\nu_{[+-,dyn]}$ & $\nu_{[+-,dyn]}^{corr}$ & $N_{part}$ &   $\nu_{[+-,dyn]}$ & $\nu_{[+-,dyn]}^{corr}$\\\hline
\multicolumn{7}{c}{\it{AuAu at 100 GeV}  errors are in units of $\times$ $10^{-3}$}  \\  \hline
5&349.19$\pm$0.10& -0.00279$\pm$0.06&-0.00134$\pm$0.67 &349.36$\pm$0.10&-0.00450$\pm$0.11&-0.00247$\pm$1.23  \\  
10&291.43$\pm$0.10& -0.00339$\pm$0.07&-0.00162$\pm$0.81 &291.54$\pm$0.10&-0.00523$\pm$0.14&-0.00278$\pm$1.39 \\ 
20&219.88$\pm$0.10& -0.00469$\pm$0.07&-0.00228$\pm$1.14 &219.95$\pm$0.10&-0.00743$\pm$0.15&-0.00418$\pm$2.09 \\ 
30&146.36$\pm$0.08& -0.00742$\pm$0.12&-0.00373$\pm$1.87 &146.21$\pm$0.08&-0.01137$\pm$0.21&-0.00647$\pm$3.24 \\   
40&91.66$\pm$0.07 & -0.01252$\pm$0.22&-0.00643$\pm$3.22 & 91.79$\pm$0.07&-0.01735$\pm$0.41&-0.00955$\pm$4.78 \\  
50&54.16$\pm$0.05 & -0.02176$\pm$0.35&-0.01126$\pm$5.63 & 54.00$\pm$0.05&-0.03048$\pm$0.56&-0.01725$\pm$8.63 \\    
60&28.81$\pm$0.04 & -0.04201$\pm$0.82&-0.02183$\pm$10.9 & 28.72$\pm$0.04&-0.05813$\pm$1.06& -0.03342$\pm$16.72\\              
70&13.73$\pm$0.02 & -0.08949$\pm$1.37&-0.04683$\pm$23.42 &13.76$\pm$0.02&-0.11431$\pm$2.79& -0.06308$\pm$31.58\\        
80&6.51$\pm$0.02 &-0.19365$\pm$5.40 &-0.10322$\pm$51.69 &6.48$\pm$0.02  &-0.23294$\pm$7.70& -0.12489$\pm$62.58 \\ \hline 
\multicolumn{7}{c}{\it{AuAu at 130 GeV}  errors are in units of $\times$ $10^{-3}$}  \\  \hline   
5&350.57$\pm$0.10&-0.00238$\pm$0.06&-0.00116$\pm$0.58 &350.56$\pm$0.10&-0.00466$\pm$0.10&-0.00276$\pm$1.38 \\
10&293.23$\pm$0.10&-0.00306$\pm$0.06&-0.00156$\pm$0.78 &293.29$\pm$0.10&-0.00531$\pm$0.16&-0.00303$\pm$1.52 \\	
20&221.67$\pm$0.10&-0.00424$\pm$0.61&-0.00219$\pm$1.14 &222.00$\pm$0.10&-0.00722$\pm$0.19&-0.00420$\pm$2.10 \\
30&147.80$\pm$0.08&-0.00666$\pm$0.77&-0.00351$\pm$1.80 &147.97$\pm$0.08&-0.01109$\pm$0.20&-0.00657$\pm$3.28 \\
40&93.12$\pm$0.07 &-0.01093$\pm$3.68&-0.00571$\pm$3.44 & 93.16$\pm$0.07&-0.01741$\pm$0.38&-0.01025$\pm$5.13 \\
50&55.14$\pm$0.05 &-0.01936$\pm$2.35&-0.01030$\pm$5.30 &55.08$\pm$0.05 &-0.02889$\pm$0.61&-0.01683$\pm$8.42 \\
60&29.49$\pm$0.04 &-0.03832$\pm$3.89&-0.02091$\pm$10.66 &50.10$\pm$0.05&-0.02884$\pm$0.70&-0.01683$\pm$8.43 \\
70&14.14$\pm$0.03 &-0.08413$\pm$8.79&-0.04713$\pm$24.07 &14.23$\pm$0.03&-0.11416$\pm$2.80&-0.06826$\pm$34.17 \\
80&6.71$\pm$0.02  &-0.17158$\pm$4.41&-0.09314$\pm$46.63 &6.74$\pm$0.02&-0.23619$\pm$7.59& -0.13987$\pm$70.08 \\ \hline
\multicolumn{7}{c}{\it{AuAu at 200 GeV}  errors are in units of $\times$ $10^{-3}$} \\  \hline       
5&353.05$\pm$0.09&-0.00216$\pm$0.05&-0.00120$\pm$0.60 &353.17$\pm$0.09&-0.00452$\pm$0.11 & -0.00279$\pm$1.39 \\
10&296.71$\pm$0.10&-0.00255$\pm$0.04&-0.00138$\pm$0.69 &296.84$\pm$0.10&-0.00544$\pm$0.14 & -0.00337$\pm$1.69 \\
20&225.37$\pm$0.10&-0.00317$\pm$0.31&-0.00160$\pm$0.81 &225.55$\pm$0.10&-0.00723$\pm$0.12 & -0.00450$\pm$2.25 \\
30&155.03$\pm$0.10&-0.00557$\pm$0.42&-0.00100$\pm$0.52 &151.28$\pm$0.08&-0.01067$\pm$0.18 & -0.00662$\pm$3.31 \\
40&95.63$\pm$0.07 &-0.00917$\pm$1.00&-0.00505$\pm$2.58 &96.00$\pm$0.07 &-0.01630$\pm$0.28 & -0.00993$\pm$4.97\\
50&56.92$\pm$0.05 &-0.01855$\pm$2.54&-0.01129$\pm$5.85 &57.39$\pm$0.05 &-0.02718$\pm$0.52 & -0.01657$\pm$8.29\\
60&31.07$\pm$0.04 &-0.03223$\pm$6.21&-0.01850$\pm$9.26 &31.16$\pm$0.04 &-0.04752$\pm$0.99 & -0.028104$\pm$14.06\\
70&14.93$\pm$0.03 &-0.06861$\pm$8.52&-0.03957$\pm$20.38 &15.12$\pm$0.03&-0.09972$\pm$2.23 & -0.059954$\pm$30.00\\
80&7.16$\pm$0.02  &-0.15018$\pm$3.64&-0.08783$\pm$43.97 &7.18 $\pm$0.02&-0.20465$\pm$6.96 & -0.121323$\pm$60.80\\ \hline \hline  
\end{tabular}
\end{table}
%+++++++++++++++++++++++++++++++++++++++++++++++++

\newpage
\begin{center}
\begin{table}
\footnotesize
\centering
\caption{Values of $N_{part}$, $\nu_{[+-,dyn]}$ and $\nu_{[+-,dyn]}^{corr}$ for different centrality bins in $|\eta| < 1.0$ simulated for $^{64}Cu - ^{64}Cu$ interactions at 62.4 and 200 GeV.}\vspace{2mm}
\begin{tabular}{c|rrr|rrr} \hline \hline
 &\multicolumn{3}{c|}{HIJING - default}  & \multicolumn{3}{c}{HIJING - nojets} \\
cent.$\%$ & $N_{part}$  &    $\nu_{[+-,dyn]}$ & $\nu_{[+-,dyn]}^{corr}$ & $N_{part}$ &   $\nu_{[+-,dyn]}$ & $\nu_{[+-,dyn]}^{corr}$\\\hline
\multicolumn{7}{c}{\it{CuCu at 62.4 GeV}  errors are in units of $\times$ $10^{-3}$}  \\  \hline 
5&103.68$\pm$0.03&-0.011595$\pm$0.26&-0.00519$\pm$2.60 &103.73$\pm$0.03&-0.01479$\pm$0.36& -0.00725$\pm$3.62 \\
10&88.36$\pm$0.04 &-0.01314$\pm$0.23&-0.00556$\pm$2.78 &88.45$\pm$0.04 &-0.01631$\pm$0.43&  -0.00743$\pm$3.72 \\
20&68.59$\pm$0.04 &-0.01813$\pm$0.31&-0.00825$\pm$4.12 &68.35$\pm$0.04 &-0.02268$\pm$0.37&  -0.01115$\pm$5.57\\
30&47.57$\pm$0.03 &-0.02577$\pm$0.44&-0.01134$\pm$5.67 &47.61$\pm$0.03 &-0.03284$\pm$0.48&  -0.01623$\pm$8.12\\ 
40&31.62$\pm$0.03 &-0.04126$\pm$0.52&-0.01937$\pm$9.69 &31.58$\pm$0.03 &-0.04972$\pm$0.75&  -0.02466$\pm$12.34 \\
50&20.45$\pm$0.02 &-0.06349$\pm$1.16&-0.02943$\pm$14.72 &20.42$\pm$0.02 &-0.07580$\pm$1.09&-0.03711$\pm$18.56\\
60&12.67$\pm$0.02 &-0.10112$\pm$2.01&-0.04593$\pm$22.98 &19.33$\pm$0.02 &-0.07352$\pm$1.31&-0.03464$\pm$17.33\\
70&7.92$\pm$0.01  &-0.16345$\pm$2.73&-0.07501$\pm$37.53 &7.91$\pm$0.01  &-0.18968$\pm$4.29&-0.09017$\pm$45.13\\
80&5.26$\pm$0.01  &-0.25073$\pm$4.09&-0.11766$\pm$58.86 &5.28$\pm$0.01  &-0.28850$\pm$6.42& -0.13992$\pm$70.03\\ \hline
\multicolumn{7}{c}{\it{CuCu at 200 GeV}  errors are in units of $\times$ $10^{-3}$}  \\  \hline
5&107.26$\pm$0.03&-0.00756$\pm$0.16&-0.00423$\pm$2.12 &107.09$\pm$0.03&-0.01468$\pm$0.45& -0.00905$\pm$4.53\\
10&92.39$\pm$0.04 &-0.00911$\pm$0.17&-0.00517$\pm$2.58 &92.31$\pm$0.04 &-0.01728$\pm$0.43& -0.01073$\pm$5.37\\
20&72.41$\pm$0.04 &-0.01174$\pm$0.18&-0.00648$\pm$3.24 &72.54$\pm$0.04 &-0.02189$\pm$0.46& -0.01355$\pm$6.78\\
30&51.26$\pm$0.03 &-0.01766$\pm$0.25&-0.00997$\pm$4.98 &51.31$\pm$0.03 &-0.03018$\pm$0.56& -0.01839$\pm$9.20\\
40&34.79$\pm$0.03 &-0.02786$\pm$0.39&-0.01613$\pm$8.06 &34.82$\pm$0.03 &-0.04548$\pm$0.96& -0.02814$\pm$14.08\\
50&22.92$\pm$0.03 &-0.04417$\pm$0.77&-0.02579$\pm$12.90 &22.88$\pm$0.03 &-0.06879$\pm$1.45&-0.04248$\pm$20.12\\
60&20.93$\pm$0.03 &-0.04410$\pm$0.64&-0.02579$\pm$12.90&20.91$\pm$0.03 &-0.06650$\pm$1.33& -0.04027$\pm$21.01\\	 
70&9.11$\pm$0.02  &-0.10984$\pm$2.26&-0.06197$\pm$31.01 &9.13$\pm$0.02 &-0.16262$\pm$4.29&  -0.09710$\pm$48.62\\
80&5.99$\pm$0.01  &-0.17367$\pm$2.84&-0.10029$\pm$50.17 &5.98$\pm$0.01 &-0.24322$\pm$6.05&  -0.14339$\pm$71.78\\ \hline \hline
\end{tabular}
\end{table}
\end{center}
%+++++++++++++++++++++++++++++++++++++++++++++++++++

\newpage
\begin{center}
\begin{table}
\footnotesize
\centering
\caption{Values of $N_{part}$, $\nu_{[+-,dyn]}$ and $\nu_{[+-,dyn]}^{corr}$ for different centrality bins in $|\eta| < 1.0$ simulated for $^{208}Pb - ^{208}Pb$ collisions at 2.76 and 5.02 TeV.}\vspace{2mm}
\begin{tabular}{c|rrr|rrr} \hline \hline
& \multicolumn{3}{c|}{HIJING - default}  & \multicolumn{3}{c}{HIJING - nojets} \\
cent.$\%$ & $N_{part}$  &    $\nu_{[+-,dyn]}$ & $\nu_{[+-,dyn]}^{corr}$ & $N_{part}$ & $\nu_{[+-,dyn]}$ & $\nu_{[+-,dyn]}^{corr}$\\\hline
\multicolumn{7}{c}{\it{PbPb at 2760 GeV}  errors are in units of $\times$ $10^{-3}$}  \\  \hline 
5&383.54$\pm$0.09&-0.00077$\pm$0.01&-0.00055$\pm$0.27 &383.75$\pm$0.09&-0.00227$\pm$0.06&-0.00160$\pm$0.80 \\
10&327.34$\pm$0.11&-0.00091$\pm$0.02&-0.00062$\pm$0.31 &333.39$\pm$1.77&-0.00279$\pm$0.06&-0.00200$\pm$1.00 \\
20&250.99$\pm$0.11&-0.00124$\pm$0.01&-0.00085$\pm$0.42 &249.97$\pm$0.11&-0.00385$\pm$0.05&-0.00282$\pm$1.41 \\
30&168.17$\pm$0.09&-0.00269$\pm$0.03&-0.00196$\pm$0.98 &172.23$\pm$0.09&-0.00517$\pm$0.10&-0.00367$\pm$1.83 \\
40&115.74$\pm$0.08&-0.00354$\pm$0.04&-0.00251$\pm$1.25 &111.91$\pm$0.08&-0.00803$\pm$0.15&-0.00571$\pm$2.85 \\
50&69.43$\pm$0.09 &-0.00640$\pm$0.14&-0.00461$\pm$2.30 &69.52$\pm$0.06 &-0.01373$\pm$0.26&-0.00997$\pm$4.99 \\
60&31.13$\pm$0.04 &-0.01060$\pm$0.62&-0.00722$\pm$3.63 &39.36$\pm$0.05 &-0.02487$\pm$0.38&-0.01815$\pm$9.08 \\
70&22.33$\pm$0.04 &-0.02685$\pm$0.31&-0.01943$\pm$9.71 &20.33$\pm$0.03 &-0.04836$\pm$0.93&-0.03506$\pm$17.54\\
80&11.25$\pm$0.04 &-0.05571$\pm$0.87&-0.03928$\pm$19.65 &9.99$\pm$0.03 &-0.10468$\pm$3.34&-0.07691$\pm$38.53 \\ \hline
\multicolumn{7}{c}{\it{PbPb at 5020 GeV}  errors are in units of $\times$ $10^{-3}$}  \\ \hline
5&385.67$\pm$0.09&-0.00056$\pm$0.01& -0.00040$\pm$0.20&386.00$\pm$0.09&-0.00225$\pm$0.06&-0.00167$\pm$0.83 \\
10&331.10$\pm$0.11&-0.00077$\pm$0.01& -0.00057$\pm$0.28&330.62$\pm$0.11&-0.00257$\pm$0.06&-0.00189$\pm$0.94 \\
20&254.11$\pm$0.11&-0.00093$\pm$0.01& -0.00064$\pm$0.32&255.20$\pm$0.11&-0.00355$\pm$0.06&-0.00267$\pm$1.33 \\
30&173.09$\pm$0.12&-0.00164$\pm$0.01& -0.00129$\pm$0.64&174.31$\pm$0.09&-0.00515$\pm$0.07&-0.00385$\pm$1.92 \\
40&116.08$\pm$0.08&-0.00277$\pm$0.03& -0.00205$\pm$1.02&116.85$\pm$0.08&-0.00755$\pm$0.13&-0.00561$\pm$2.80 \\
50&73.34$\pm$0.06 &-0.00461$\pm$0.06& -0.00339$\pm$1.69&72.65$\pm$0.06 &-0.01261$\pm$0.22&-0.00947$\pm$4.73\\
60&42.17$\pm$0.05 &-0.00941$\pm$0.12& -0.00702$\pm$3.51 &42.02$\pm$0.05&-0.02167$\pm$0.43&-0.01617$\pm$8.09 \\
70&22.33$\pm$0.04 &-0.01991$\pm$0.31 &-0.01482$\pm$7.41 &22.13$\pm$0.04&-0.04158$\pm$1.20&-0.03094$\pm$15.49 \\
80&11.25$\pm$0.04 &-0.04568$\pm$0.87 &-0.03465$\pm$17.34 &11.05$\pm$0.03&-0.08586$\pm$3.72&-0.0639$\pm$32.10 \\ \hline \hline
\end{tabular}
\end{table}
\end{center}

\newpage
\begin{center}
\begin{table}
\footnotesize
\centering
\caption{Values of $N_{part}$, $\nu_{[+-,dyn]}$ and $\nu_{[+-,dyn]}^{corr}$ for $^{208}Pb - ^{208}Pb$ collisions at 2.76 TeV [Data from ref. 20]}\vspace{2mm}
\begin{tabular}{c|ccc} \hline 
cent.$\%$ & $N_{part}$  &    $\nu_{[+-,dyn]}$ & $\nu_{[+-,dyn]}^{corr}$ \\\hline
5 &382.80$\pm$3.1&-0.00104$\pm$0.00001&-0.00093$\pm$0.00001 \\
10&329.70$\pm$4.6&-0.00126$\pm$0.00001&-0.00113$\pm$0.00002 \\
20&260.50$\pm$4.4&-0.00165$\pm$0.00001&-0.00148$\pm$0.00001 \\
30&186.40$\pm$3.9&-0.00236$\pm$0.00001&-0.00211$\pm$0.00002 \\
40&128.90$\pm$3.3&-0.00348$\pm$0.00008&-0.00311$\pm$0.00008 \\
50&85.00$\pm$2.6 &-0.00541$\pm$0.00004&-0.00483$\pm$0.00004  \\
60&52.80$\pm$2.0 &-0.00903$\pm$0.00007&-0.00802$\pm$0.00007 \\
70&30.00$\pm$2.8 &-0.01675$\pm$0.00017&-0.01482$\pm$0.00017 \\
80&15.80$\pm$3.8 &-0.03547$\pm$0.00041&-0.03144$\pm$0.00041 \\\hline
\end{tabular}
\end{table}
\end{center}
%====================Figures==================================================

\newpage
\begin{figure} []
\begin{center}\mbox{\psfig{file=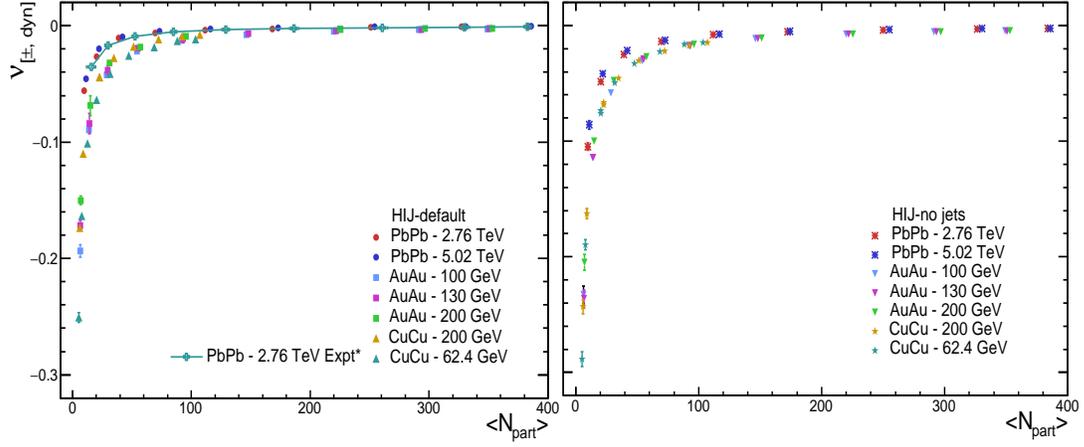,width=\textwidth,height=6cm}}
\end{center}
\caption [Fig1.]{\sf{ Dependence of net charge fluctuations, $\nu_{[+-,dyn]}$ on the number of participating nucleons, $N_{part}$ for the \hij events with jets/minijets on and off. Experimental results for Pb-Pb collisions at 2.76 TeV are also shown [Data ref 20.]}}
\label{lindat}
\end{figure} 

\newpage
\begin{figure} []
\begin{center}\mbox{\psfig{file=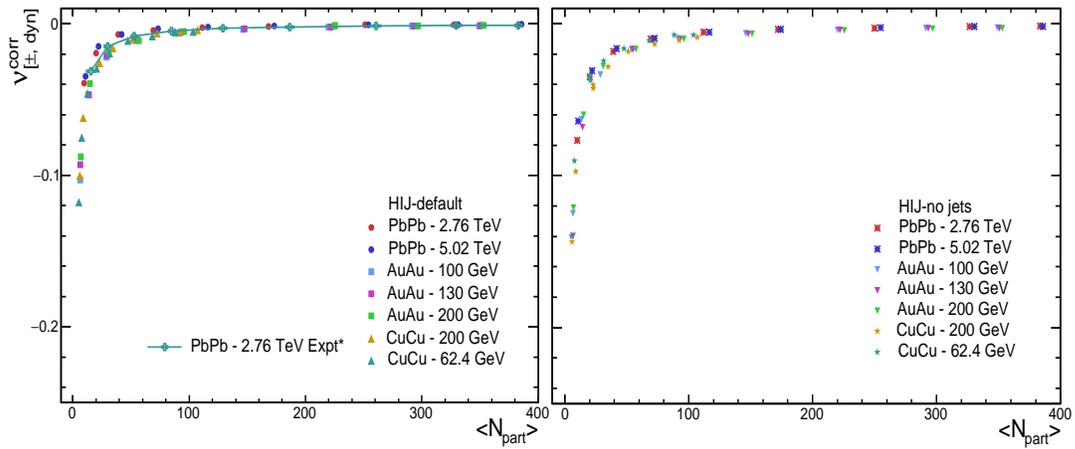,width=\textwidth,height=6cm}}
\end{center}
\caption [Fig2.]{\sf{The same plot as Fig.1 but for corrected versions of net charge fluctuations.}}
\label{lindat}
\end{figure} 

\newpage
\begin{figure} []
\begin{center}\mbox{\psfig{file=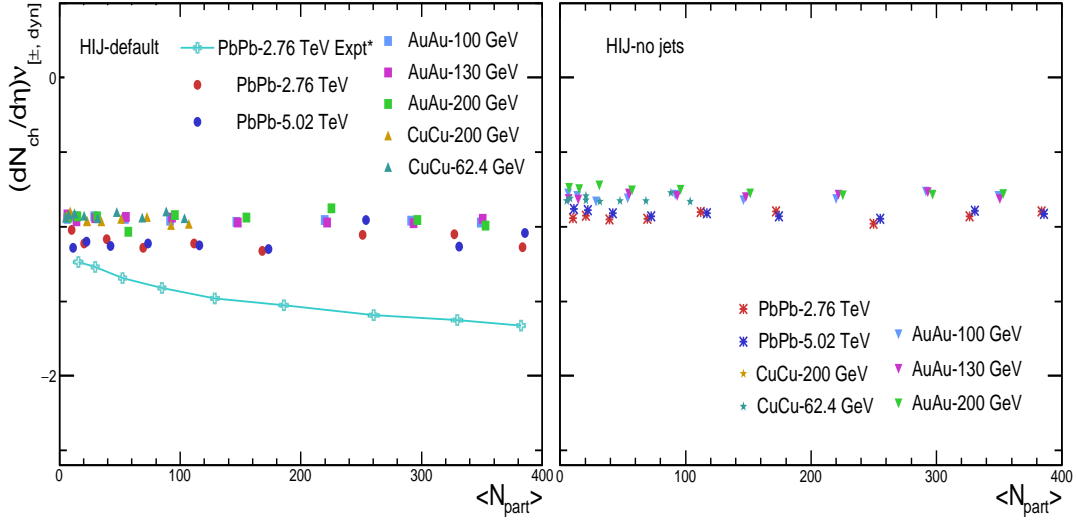,width=\textwidth,height=7cm}}
\end{center}
\caption [Fig3.]{\sf{$(dN_{ch}/d\eta)\nu_{[+-,dyn]}$ plotted against $N_{part}$ for \hij default with jet production on(left panel) and jet production off(right panel). The line represents the Pb-Pb data from ref 20.}}
\label{lindat}
\end{figure} 

\newpage
\begin{figure} []
\begin{center}\mbox{\psfig{file=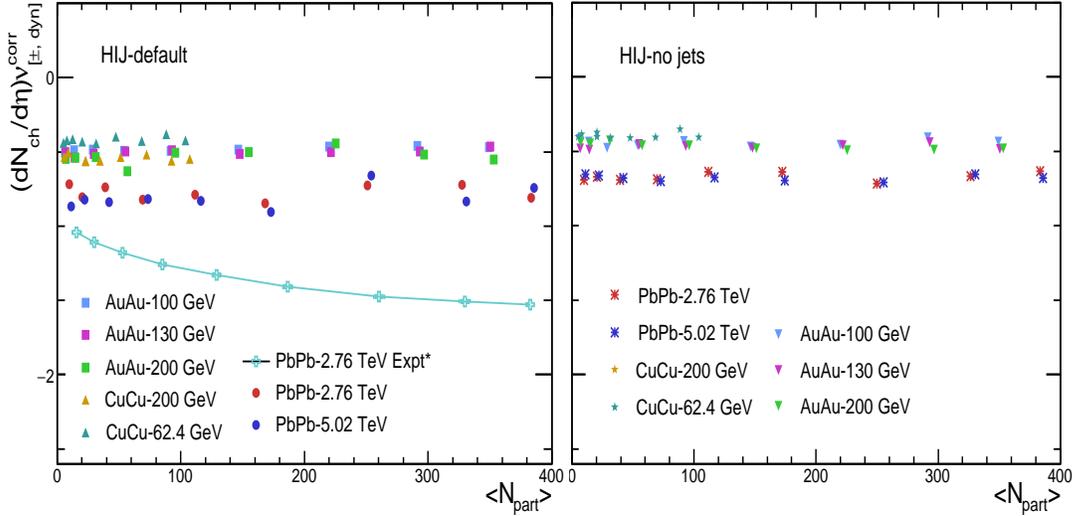,width=\textwidth,height=7cm}}
\end{center}
\caption [Fig4.]{\sf{The same plot as Fig.3 but for corrected net charge fluctuations, $\nu_{[+-,dyn]}^{corr}$}}
\label{lindat}
\end{figure}

\newpage
\begin{figure} []
\begin{center}\mbox{\psfig{file=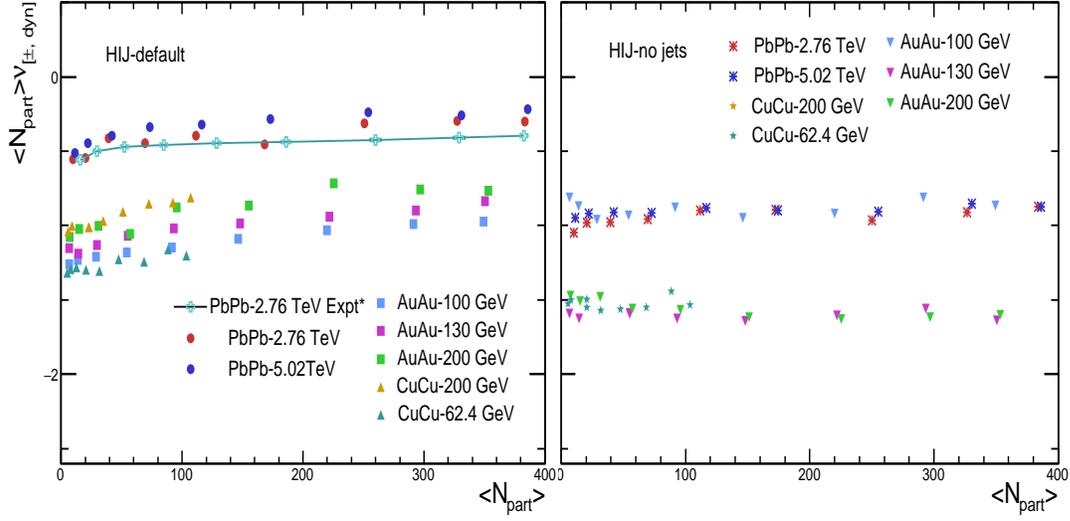,width=\textwidth,height=7cm}}
\end{center}
\caption [Fig5.]{\sf Dependence of product of $N_{part}$ and $\nu_{[+-,dyn]}$ on centrality for the two sets of \hij events at different energies. The line represents the experimental result reported in ref 20 for $\sqrt{s_{NN}} =$ 2.76 TeV Pb-Pb collisions.}
\label{lindat}
\end{figure} 

\newpage
\begin{figure} []
\begin{center}\mbox{\psfig{file=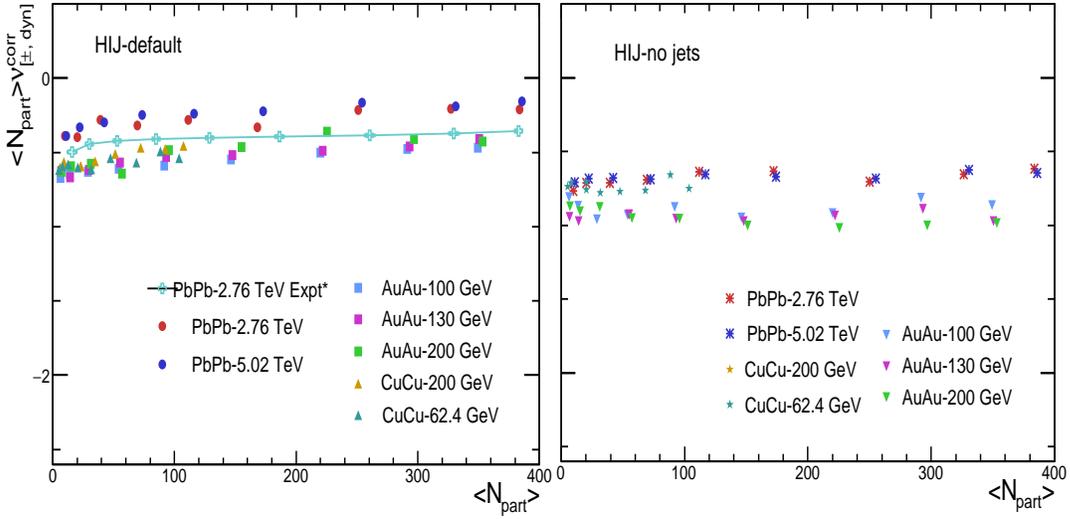,width=\textwidth,height=7cm}}
\end{center}
\caption [Fig6.]{\sf Variations of $(N_{part})\nu_{[+-,dyn]}^{corr}$ with $N_{part}$ for the two sets of \hij events.}
\label{lindat}
\end{figure} 

\newpage
\begin{figure} []
\begin{center}\mbox{\psfig{file=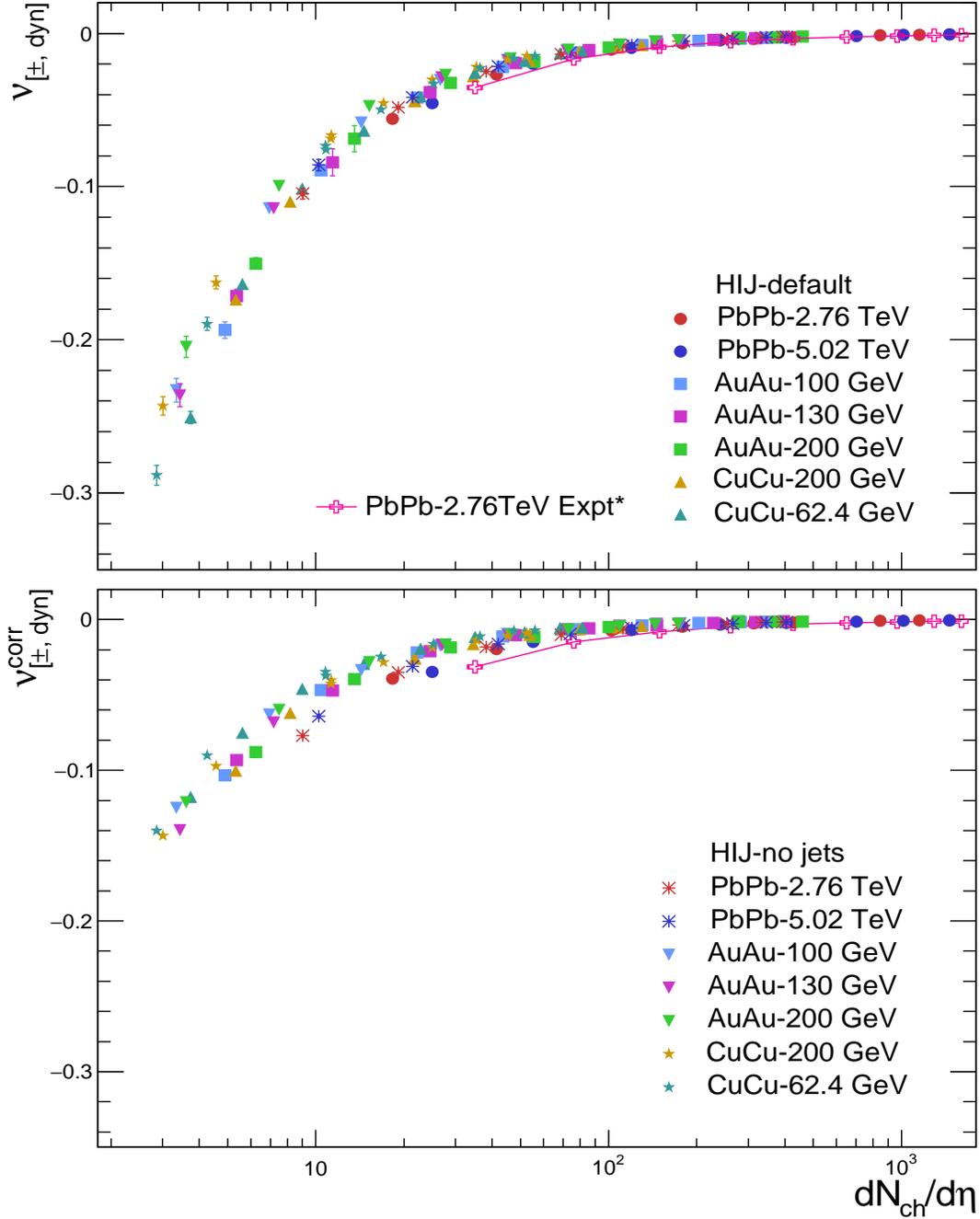,width=\textwidth,height=18cm}}
\end{center}
\caption [Fig7.]{{\sf Variations of net charge fluctuations, $\nu_{[+-,dyn]}$ and its corrected version, $\nu_{[+-,dyn]}^{corr}$ with charged particle density, $dN_{ch}/d\eta$ for the two sets of \hij events. The lines are due to the 2.76 TeV Pb-Pb values taken from ref 20.}}
\label{lindat}
\end{figure}

\newpage
\begin{figure} []
\begin{center}\mbox{\psfig{file=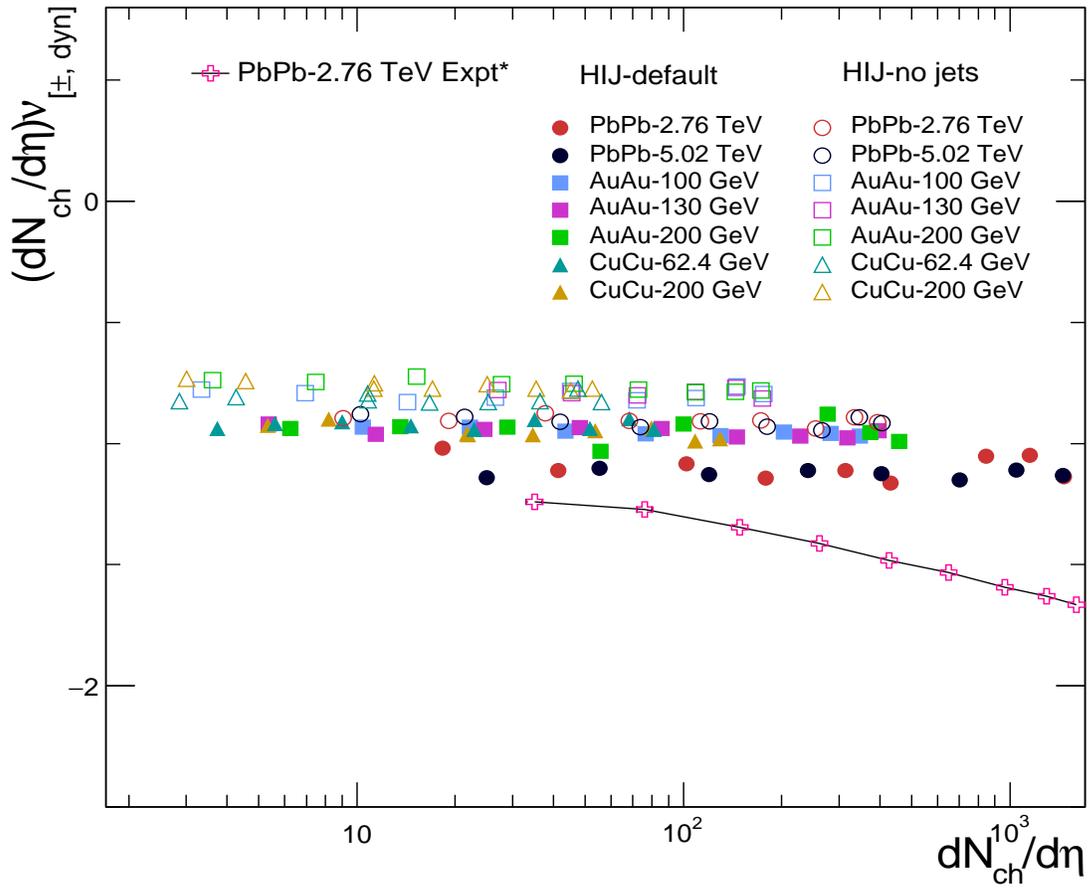,width=\textwidth,height=12cm}}
\end{center}
\caption [Fig8.]{{\sf Scaling of $\nu_{[+-,dyn]}$ with $dN_{ch}/d\eta$ for various MC data samples at different energies.}}
\label{lindat}
\end{figure} 

\newpage
\begin{figure} []
\begin{center}\mbox{\psfig{file=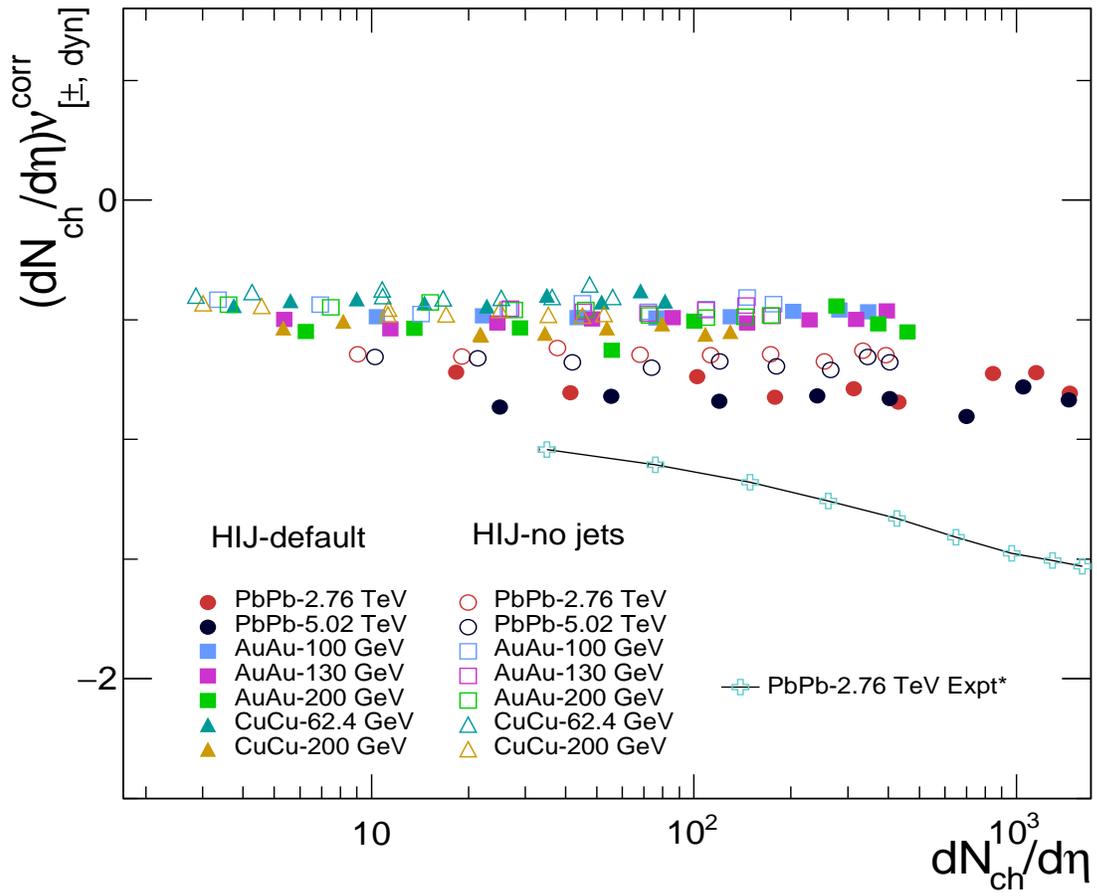,width=\textwidth,height=12cm}}
\end{center}
\caption [Fig9.]{{\sf The same plot as in Fig.8 but after applying corrections to $\nu_{[+-,dyn]}$ values.}}
\label{lindat}
\end{figure} 

\newpage
\begin{figure} []
\begin{center}\mbox{\psfig{file=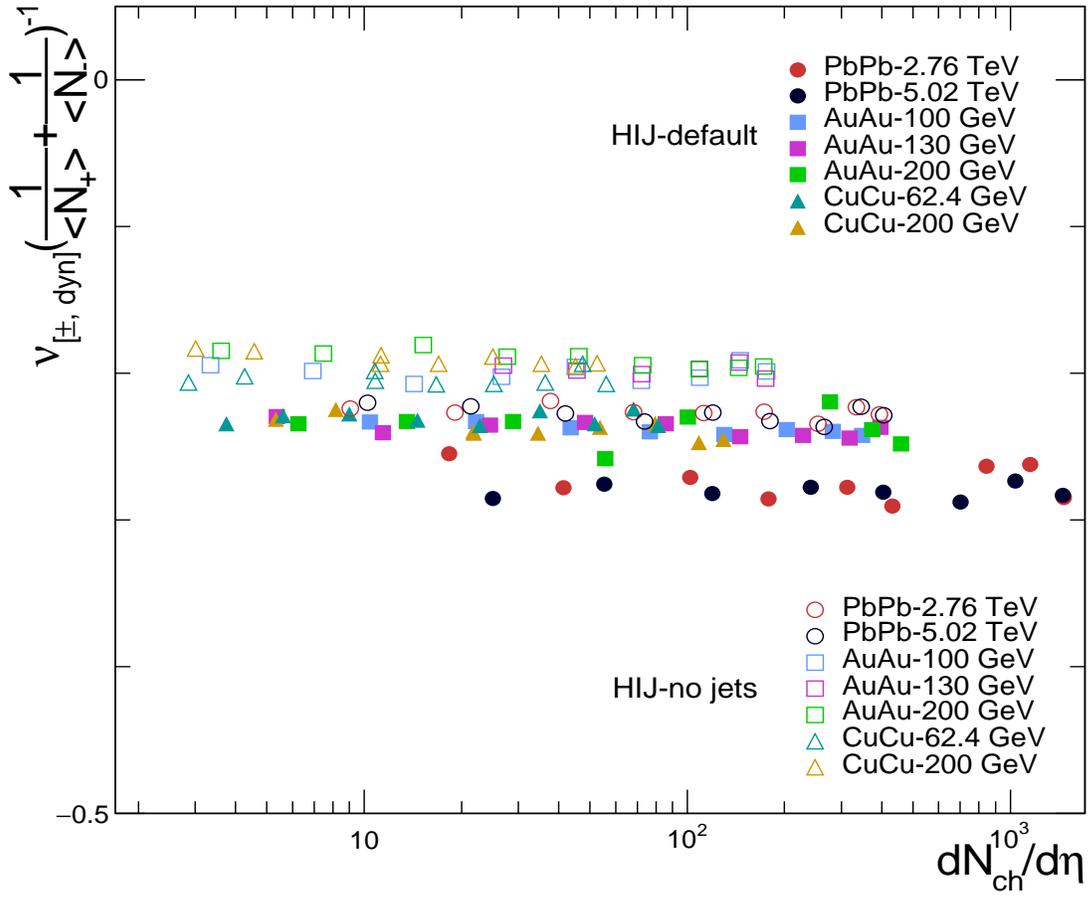,width=\textwidth,height=12cm}}
\end{center}
\caption [Fig10.]{{\sf  $1/(\frac{1}{\langle N_+\rangle}+\frac{1}{\langle N_-\rangle})$ scaling of net charge fluctuations at different energies for the two sets of \hij events.}}
\label{lindat}
\end{figure}

\newpage
\begin{figure} []
\begin{center}\mbox{\psfig{file=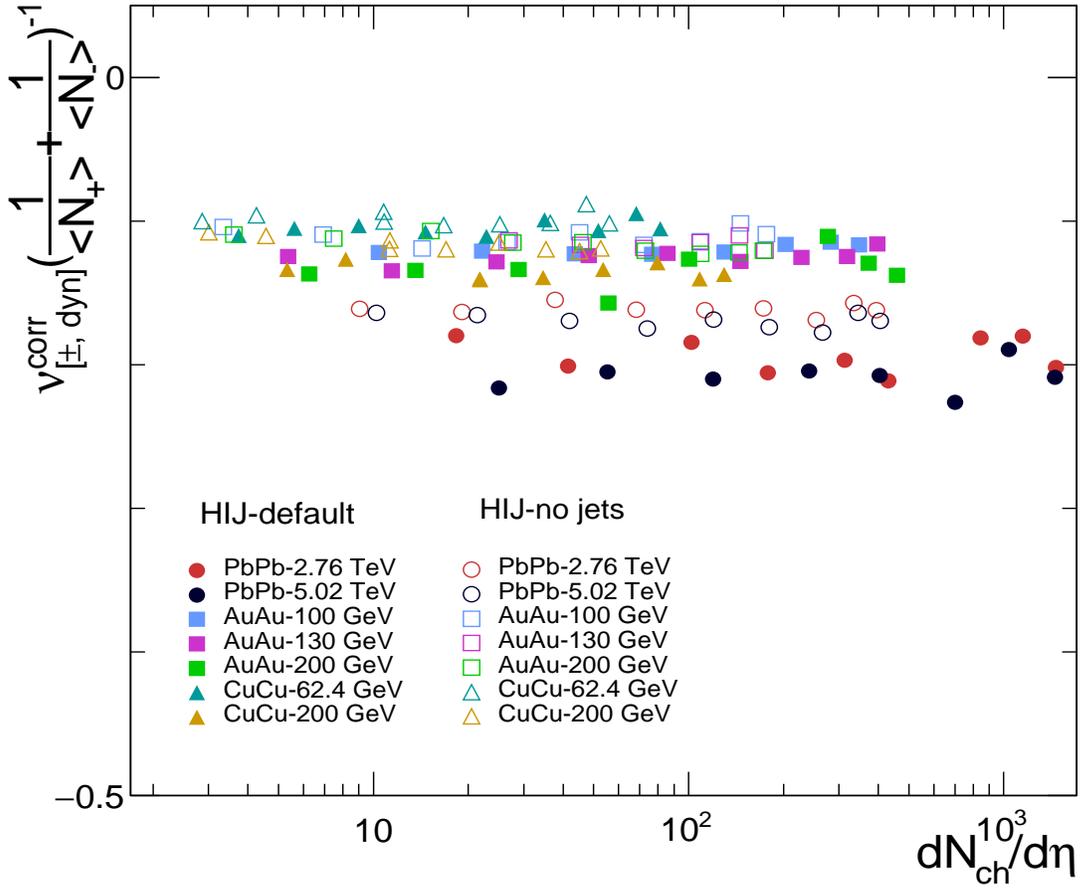,width=\textwidth,height=12cm}}
\end{center}
\caption [Fig11.]{{\sf  $1/(\frac{1}{\langle N_+\rangle}+\frac{1}{\langle N_-\rangle})$ scaling of corrected net charge fluctuations for two types of \hij events at different energies.}}
\label{lindat}
\end{figure}

\end{document}